\documentstyle[titlepage,fleqn,11pt]{article}
\begin{document}
%\rightline{IFUG-95-RSO-1}
%\rightline{September 30, 1995}
\rightline{cond-mat/9510016}
\rightline{J. Phys. A: Math. Gen. {\bf 29}, 1825-1829 (1996)}
\begin{center}{\Large{\bf Two-dimensional
Fokker-Planck solutions and Grassmann variables}}\\
\vskip 3mm
H C Rosu$^{\dagger}$$^{\ddagger}$\footnote{e-mail: rosu@ifug.ugto.mx},
J Socorro$^{\dagger}$\footnote{e-mail: socorro@ifug.ugto.mx} and
O Obreg\'on$^{\dagger}$$^{\P}$\footnote{e-mail: octavio@ifug.ugto.mx}\\

$^{\dagger}$
Instituto de F\'{\i}sica de la Universidad de Guanajuato, Apdo Postal
E-143, L\'eon, Gto, M\'exico\\

$^{\ddagger}$
Institute of Gravitation and Space Sciences, PO Box MG-6, Magurele-Bucharest,
Romania\\

$^{\P}$
Universidad Aut\'onoma Metropolitana, Apdo Postal 55-534, M\'exico Distrito
Federal, M\'exico

\vskip 0.3cm

{\bf Abstract}

After a short outline of the factorization and Grassmann picture of
the one-dimensional (1D) Fokker-Planck (FP) equation, we consider a class of
spatially-inhomogeneous solutions of the 2D FP equation
with symmetric 2D (super)potentials. We show that the spatial
inhomogeneities of that class of solutions
can be attributed to underlying Grassmannian pseudo-degrees
of freedom. Such an interpretation may also be applied to FP solutions in
three and more dimensions.

\end{center}

\bigskip

\bigskip
PACS number(s): 05.40.+j, 11.30.Pb, 02.50.Ey, 05.20.-y

\newpage
%%%%%%%%%%%%%%%%%%%%%%%%%%%  THE PAPER  %%%%%%%%%%%%%%%%%%%%%%%%%%%%%%%%
                  %%%   written by H.C. Rosu  %%%
                    %   August-September 1995 %
%%%%%%%%%%%%%%%%%%%%%%%%%%%%%%%%%%%%%%%%%%%%%%%%%%%%%%%%%%%%%%%%%%%%%%%%

\section{Introduction}
%%%%%%%%%%%%%%%%%%%%%%%%%%%%%%%%%%%%%%%%%%%%%%%%%%%%%%%%%%%%%%%%%%%%%%%

Supersymmetric techniques are by now widely used in a rich spectrum of
physical problems, covering as diverse fields as quantum gravity,
quantum cosmology,
particle physics, quantum field theory and statistical mechanics.
In the latter area, Grassmann variables have been first
considered by
Parisi and Sourlas \cite{ps} when they have cast into supersymmetric form a
simple model of a scalar field coupled to a random external source.
These intriguing variables have been
most often used for Langevin equations with complex actions (Langevin
equation formulation of quantum field theory \cite{f}), but lately their
application have been considerably extended. As remarked by
Sourlas \cite{s} the anticommuting variables are introduced more for
combinatorial and other technical reasons and do not correspond to
spin degrees of freedom as one might naively think.
On the other hand, in the context of
the Brownian diffusion in a 1D bistable potential \cite{susy} the simple Witten
supersymmetric procedure \cite{W} has led
to a remarkable elegant way of computing the smallest nonvanishing eigenvalue
$\lambda _1$, which is known to characterize the relaxation rate
towards equilibrium of a stochastic system.

In this work, our aim is to show that a class of solutions of the
2D FP equation with symmetric 2D superpotential may be
interpreted in terms of an underlying Grassmannian structure. The formalism
is essentially that employed in quantum supercosmology \cite{qc},
where however, as a rule, the cosmological potential is given and one
works out the corresponding superpotential. The FP situation is exactly
the opposite one. The formalism can be easily generalized to more
coordinates.

The organization of the paper is as follows. In the next section we
outline Witten's scheme \cite{W} for the 1D FP equation \cite{susy}
including also the superspace extension. In Sec. 3 we present a
class of 2D FP solutions that can be traced to
Grassmannian pseudo degrees of freedom and we end up with some conclusions.

%%%%%%%%%%%%%%%%%%%%%%%%%%%%%%%%%%%%%%%%%%%%%%%%%%%%%%%%%%%%%%%%%%%%%%%%%
\section{Factorization of the 1D FP operator}
%%%%%%%%%%%%%%%%%%%%%%%%%%%%%%%%%%%%%%%%%%%%%%%%%%%%%%%%%%%%%%%%%%%%%%%%%%

Over a decade ago, Bernstein and Brown provided a simple discussion of the
correspondence between the 1D FP equation with an
arbitrary potential and Witten's supersymmetric quantum mechanics.
The 1D FP equation with constant diffusion coefficient (here normalized to
unity) and potential drift is

$$\frac{1}{\gamma}\frac{\partial}{\partial t}{\cal P }(x,t)=
\frac{1}{\sqrt{\gamma}}\frac{\partial}{\partial x}
\Bigg[\frac{1}{\sqrt{\gamma}}\frac{\partial}{\partial x}+
\sqrt{\gamma} U'(x)\Bigg]{\cal P}(x,t)
\eqno(1)
$$
%%%%%%%%%%%%%%%%%%%%
where $U'=dU/dx$ is the drift force up to a sign, and $\gamma= 1/k_BT$ in
the case of the approach to thermal
equilibrium \cite{susy}, whilst $\gamma=1/\nu m$, ($\nu$ being the collision
frequency and $m$ the molecular mass), in chemical reaction kinetics
for a system with two stable states, e.g., the {\em trans-gauche}
izomerization process \cite{tg}. For more algebraic symmetry, we introduced an
overall $\frac{1}{\gamma}$ factor in equation (1), which can always be cast
into the Schr\"odinger equation as follows.
Any initial time-dependent distribution $P(x,t)$ will relax at asymptotic
times to the static solution

$${\cal P}_{st}(x)=const \exp [-\gamma U(x)]
\eqno(2)
$$
%%%%%%%%%%%%%%%
where $const$ is a normalization constant. The evolution at intermediate
times can be discussed conveniently by means of the celebrated ansatz
$${\cal P}(x,t)= \varphi (x,t) \exp \left(-\frac{1}{2}\gamma U(x)\right)~.
\eqno(3)
$$
%%%%%%%%%%%%%%%%%
${\cal P}(x,t)\rightarrow {\cal P}_{st}$
when $t\rightarrow \infty$. It turns the FP evolution of ${\cal P}$ into
a Schr\"odinger evolution for $\varphi$ in imaginary time
$\frac{1}{\gamma}\frac{\partial \varphi}{\partial t}=-H_{FP}\varphi$,
where the FP Hamiltonian is a Hermitian and positive semidefinite operator.
It is now easy to proceed with the factorization and the
whole Witten scheme. We write
$ H_{FP,1}=A^{\dagger}A
$
with
$A=\frac{1}{\sqrt{\gamma}}
\frac{\partial}{\partial x} - \sqrt{\gamma} U'$
and
$A^{\dagger}=\frac{1}{\sqrt{\gamma}}
\frac{\partial}{\partial x} + \sqrt{\gamma} U'$.
Thus, the FP superpotential is proportional to the drift force, and
the superpartner Hamiltonian will be $H_{FP,2}= AA^{\dagger}$. The two FP
Hamiltonian partners are defined as usual
$$-H_{FP,1,2}=\frac{1}{\gamma}\frac{d^2}{dx^2}+V_{1,2}
\eqno(4)
$$
%%%%%%%%%%%%%%%%%%%%%%%
with the potentials $V_{1,2}$ entering simple Riccati equations
$V_{2,1}=-\gamma U'^{2}\pm U''$. As it is
well-known
the great advantage of the supersymmetric procedure for the FP problem
is to replace bistable `bosonic' potentials with much simpler
single-well `fermionic' ones \cite{susy}.
The equilibrium distribution can always be written in terms of the FP
superpotential as
$${\cal P}_{st}(x)= const \exp \left(-2\int W_{FP}(x)dx\right)~.
\eqno(5)
$$
%%%%%%%%%%%%%%%%%%%%%%%%
The time dependence of $\varphi$ can be exponentiated $\varphi (x,t)=
\varphi (x)\exp(-\lambda t)$ \cite{exp} leading to the stationary Schr\"odinger
equation
$$
\frac{1}{\gamma}
\frac{d^2\varphi (x)}{dx^2}+[\lambda+V_{1,2}(x)]\varphi (x)=0~.
\eqno(6)
$$
%%%%%%%%%%%%%%%%%%%%

In the following we shall pay particular attention to the spatial function
$\varphi (x)$. We would like to exploit some features of the formalism that
may come out when considering it as a superfield.
We shall make use of some simple rules of the Grassmannian calculus as given
in Berezin \cite{ber}.

To introduce the superspace extension of the 1D FP equation,
one should write the FP supercharges as follows
$$Q=\psi\Bigg[-i\frac{1}{\sqrt{\gamma}}
\frac{\partial}{\partial x}+i\sqrt{\gamma}\frac{\partial U}{\partial x}\Bigg ]
\eqno(7a) $$
%%%%%%%%%%%%%%%%%%%%%
and
$$\bar {Q}=\bar \psi\Bigg[-i\frac{1}{\sqrt{\gamma}}\frac{\partial}{\partial x}-
i\sqrt{\gamma}\frac{\partial U}{\partial x}\Bigg]~.
\eqno(7b)$$
%%%%%%%%%%%%%%%%%%%%%
where $\psi=\frac{\partial}{\partial \theta ^0}$ and $\bar \psi=\theta ^0$
in the Grassmann representation. The superspace 1D FP Hamiltonian operator
will be
$$
H_{FP}=\{Q,\bar Q\}=\Bigg[P^2+\left(\frac{\partial U}{\partial x}\right)^2 -
\frac{\partial ^2 U}{\partial x^2}\Bigg]+ \theta ^0
\frac{\partial}{\partial \theta ^0}\frac{\partial ^2 U}{\partial x^2}
\eqno(8)
$$
%%%%%%%%%%%%%%%%%%%%%%%
where $P=i\frac{1}{\sqrt{\gamma}}\frac{\partial}{\partial x}$ is the FP
momentum operator.

Since we are interested in possible spatial inhomogeneities of
``fermionic"-Grassmann origin
we shall expand the $\varphi$-superfield in Grassmann variables as follows
$$\varphi(x,\theta ^0)= A_{+}(x) + B_{0}(x)\theta ^{0}~.
\eqno(9) $$
%%%%%%%%%%%%%%%%%%%%%%%%
The conditions $Q\varphi=0$ and $\bar{Q}\varphi=0$ defining the `ground-state'
give
$A_+=a_+\exp(-\gamma U)$ and $B_0=b_0\exp(\gamma U)$. In other words, $A_+$ is
required to be `square-integrable' in superspace,
and therefore $U(x)\rightarrow \infty$ as
$x\rightarrow \pm \infty$. Thus, one can see that the $B_{0}$
component coming from the Grassmann variable can be discarded on physical
reasons.
However, had we considered two, or more Grassmann variables the conclusion
would have been different, as it will be seen in the next section.

%%%%%%%%%%%%%%%%%%%%%%%%%%%%%%%%%%%%%%%%%%%%%%%%%%%%%%%%%%%%%%%%%%%%%%%%%
\section{2D FP solutions and Grassmann variables}
%%%%%%%%%%%%%%%%%%%%%%%%%%%%%%%%%%%%%%%%%%%%%%%%%%%%%%%%%%%%%%%%%%%%%%%%

We now show that there exist solutions of the 2D
FP equation with spatial components that can be naturally
attributed to Grassmannian variables. Consider the following probability
$$
{\cal P}= [a_+^2 +g_0^2(x)+g_1^2(y)]\exp(-2\gamma U)+
[a_-^2 ]\exp(+2\gamma U)
\eqno(10)
$$
%%%%%%%%%%%%%%%%%%%%%
where $g_0$ and $g_1$ are arbitrary functions of $x$ and $y$, respectively.
Suppose ${\cal P}$ is a solution of a 2D FP equation. Then
the term with the positive exponent is discarded as not physical. The
solution given by equation (10) can be also interpreted as a solution of the
2D FP equation in superspace.

In the 2D case, the supercharges read
$$
Q=\psi ^{\mu}\Bigg[-P_{\mu}+i\sqrt{\gamma}\frac{\partial U}{\partial q_{\mu}}
\Bigg]
\eqno(11a)
$$
%%%%%%%%%%%%%%%%%%%%%%%%
and
$$
\bar Q=\bar\psi ^{\nu}\Bigg[-P_{\nu}-i\sqrt{\gamma}
\frac{\partial U}{\partial q_{\nu}}
\Bigg]~.
\eqno(11b)
$$
%%%%%%%%%%%%%%%%%%%%%%%%
The FP momentum operators are $P_{\mu}=i\frac{1}{\sqrt{\gamma}}
\frac{\partial}{\partial q_{\mu}}$.
With $\{ \psi^{\mu}, \bar\psi^{\nu} \}$=$\eta ^{\mu \nu}$,
where the metric is
$\eta ^{\mu\nu}$=diag (1,1), i.e., $\psi^{\mu}$=$\eta^{\mu\nu}
\frac{\partial}{\partial \theta _{\nu}}$ and $\bar\psi^{\nu}=\theta^{\nu}$, one
will find the superspace FP Hamiltonian to be written in the form
$$H_{FP}=\{Q,\bar Q\} =
\eta^{\mu\nu}\Bigg[ P_{\mu}P_{\nu}+\frac{\partial U}{\partial q_{\mu}}
\frac{\partial U}{\partial q_{\nu}}-
\frac{\partial ^2 U}{\partial q_{\mu}\partial q_{\nu}}\Bigg]+
\bar\psi^{\nu}\psi^{\mu}
\frac{\partial ^2U}{\partial q_{\mu}\partial q_{\nu}}~.
\eqno(12)
$$
%%%%%%%%%%%%%%%%%%%%%%%%%
The 2D Grassmann representation of the $\varphi$ field reads
$$
\varphi= A_+ +B_0\theta ^0 +B_1\theta ^1 +A_-\theta ^0\theta ^1~.
\eqno(13)
$$
%%%%%%%%%%%%%%%%%%%%%%%
The ground-state amplitudes are determined by the conditions
$Q\varphi=0$ and $\bar{Q}\varphi=0$ with the Qs and $\varphi$ substituted
from equations (11a,b) and equation (13), respectively.  Using the
ansatz \cite{qc}
$B_{\mu}=\frac{\partial f}{\partial q_{\mu}}\exp(-\gamma U)$, $\mu$=0,1,
these conditions lead
to the following equation for $f$
$$\Box _2 f-2\gamma\nabla _2U\nabla _2f=0
\eqno(14)
$$
%%%%%%%%%%%%%%%%%%%%%%%%%
where
$$
\Box _2=\eta ^{\mu\nu}P_{\mu}P_{\nu}
\eqno(15)
$$
%%%%%%%%%%%%%%%%%%%%%%%%%
and
$$
\nabla _2U\nabla _2f=
\eta ^{\mu\nu}\frac{\partial U}{\partial q_{\mu}}
\frac{\partial f}{\partial q_{\nu}}
\eqno(16)
$$
%%%%%%%%%%%%%%%%%%%%%%%%%%

In the case of separable potential functions $U(x,y)=U_0(x)+U_1(y)$ and
separable $f$ functions, $f=f_0(x)+f_1(y)$ one will obtain equation (14)
in the form
$$
\frac{\partial ^2f_0}{\partial x^2}-2\gamma\frac{\partial U_0}{\partial x}
\frac{\partial f_0}{\partial x}= -\left(\frac{\partial ^2 f_1}{\partial y^2}
-2\gamma \frac{\partial U_1}{\partial y}\frac{\partial f_1}{\partial y}\right)
\eqno(17)
$$
%%%%%%%%%%%%%%%%%%%%%%%
which is in almost separable form. Denoting the $f$-derivatives by $F_{0}$ and
$F_{1}$, and the $U$-derivatives by $V_0$ and $V_1$ one will get
$$
\frac{\partial F_0}{\partial x}-2\gamma V_0F_0=a_1
\eqno(18a)
$$
%%%%%%%%%%%%%%%%%%%%%%
and
$$
\frac{\partial F_1}{\partial y}-2\gamma V_1F_1=-a_1~,
\eqno(18b)
$$
%%%%%%%%%%%%%%%%%%%%%%
where $a_1$ is the separation constant.
For $a_1$ = 0, one gets easily $F_0=const\exp(2\gamma U_0)$
and $F_1=const\exp(2\gamma U_1)$. The solutions for $B_0$ and $B_1$ come out
as follows $B_0=b_0\exp[\gamma(U_0-U_1)]$ and $B_1=b_1\exp[-\gamma(U_0-U_1)]$,
where $b_0$ and $b_1$ are constants.
Thus, the probability will be the square of the probability amplitude
in superspace \cite{ber}
$$
{\cal P}=[a_+^2 +b_0^2\exp(4\gamma U_0)+b_1^2\exp(4\gamma U_1)]
\exp(-2\gamma U) + a_-^2 \exp(2\gamma U)
\eqno(19)
$$
%%%%%%%%%%%%%%%%%%%%%%%
which corresponds to equation (10) above.

When the separation constant is different from zero, one should write
down the solutions of equations (18a,b), which are as follows
$$
F_0=a_1\exp(2\gamma U_0)\int\exp(-2\gamma U_0)dx +b_0\exp(2\gamma U_0)
\eqno(20a)
$$
%%%%%%%%%%%%%%%%%%%%%%%%
and
$$
F_1=-a_1\exp(2 \gamma U_1)\int\exp(-2\gamma U_1)dy +b_1\exp(2\gamma U_1)
\eqno(20b)
$$
%%%%%%%%%%%%%%%%%%%%%%%%%%
from which the $B$ coefficients are immediately obtained by multiplying
with $\exp(-\gamma U)$. When $a_1=0$ we recover the previous results.

%%%%%%%%%%%%%%%%%%%%%%%%%%%%%%%%%%%%%%%%%%%%%%%%%%%%%%%%%%%%%%%%%%%%%%%%
\section{Conclusions}
%%%%%%%%%%%%%%%%%%%%%%%%%%%%%%%%%%%%%%%%%%%%%%%%%%%%%%%%%%%%%%%%%%%%%%%%

We have shown, on the base of a simple separable example, that 2D FP
spatially-inhomogeneous solutions
can be related to Grassmannian pseudo-degrees of freedom. Therefore one
might think of them not only as a technical detail but also as having direct
physical effects.
One can easily build higher-dimensional superfields corresponding to even
more complicated spatial inhomogeneities of either non-equilibrium
thermodynamics or chemical kinetics.
In 3D, equation (13) is written as follows
$$
\varphi=A_++B_{\nu}\theta ^{\nu}+\frac{1}{2}\epsilon _{\mu\nu\lambda}
C^{\lambda}\theta ^{\mu}\theta ^{\nu} +A_-\theta ^{0}\theta ^{1}\theta ^{2}
\eqno(21)
$$
%%%%%%%%%%%%%%%%%%%%%%%%%%%%%%
where $\mu,\nu$=0,1,2,
and $B_{\mu}=\frac{\partial f}{\partial q_{\mu}}\exp(-\gamma U)$,
$C_{\lambda}=\frac{\partial f}{\partial q_{\lambda}}\exp (\gamma U)$ and
the Euclidian metric is $\eta ^{\mu \nu}=(1,1,1)$.
equation (14) is transformed into
$\Box _3 f\pm2\gamma\nabla _3U\nabla _3f=0$,
where the operators have the same meaning as in equations (15) and (16),
except that the 3D metric is used.
Equations (11) and (12) are preserved in form but the
attached $\eta$-metric is 3D.

%%%%%%%%%%%%%%%%%%%%%%%%%  END OF THE PAPER  %%%%%%%%%%%%%%%%%%%%%%%%%%%%%%
\section*{Acknowledgments}
We thank CONACyT for partial financial support through CONACyT Projects
4862-E9406 and 4868-E9406.

\newpage
%%%%%%%%%%%%%%%%%%%%%%%%%  BIBLIOGRAPHY  %%%%%%%%%%%%%%%%%%%%%%%%%%%%%%%%%%%%

\end{document}